\documentclass[letterpaper,twocolumn,10pt]{article}
\usepackage{usenix2019_v3}

\usepackage{tikz}
\usepackage[utf8]{inputenc}
\usepackage{array,multirow}
\usepackage{xspace}
\usepackage{color, colortbl}
\usepackage{balance}
\usepackage{paralist}
\usepackage{amsfonts,amssymb}
\usepackage{amsthm,amsmath}
\usepackage{hyperref}
\usepackage{cleveref}
\usepackage{graphicx}
\usepackage{pifont}
\usepackage{array}
\usepackage{booktabs}
\usepackage{tikz}
\usepackage{bm}
\usepackage[inline]{enumitem}
\usepackage{todonotes}
\usepackage{arydshln}
\usepackage{makecell}
\usepackage{tabularx}
\usepackage[usestackEOL]{stackengine}
\usepackage{caption}
\usepackage{subcaption}
\usepackage{soul}
\usepackage{ifthen}

\newboolean{showcomments}
\setboolean{showcomments}{true}
\ifthenelse{\boolean{showcomments}}
{ \newcommand{\mynote}[3]{
    \protect\fbox{\bfseries\sffamily\scriptsize#1}
    {\small$\blacktriangleright$\textsf{\emph{\color{#3}{#2}}}$\blacktriangleleft$}}}
{ \newcommand{\mynote}[3]{}}

\newcommand\sysname{EL~PASSO\xspace}

\newcommand\elgamal{El-Gamal\xspace}

\newcommand\ps{PS Signatures\xspace}

\newcommand\hashtopoint{H^*\xspace}

\newcommand{\eg}{\textit{e.g.}\@\xspace}
\newcommand{\ie}{\textit{i.e.}\@\xspace}

\newcommand\para[1]{\vspace{0.05in} \noindent \textbf{#1.}}

\newcommand\newdef[2]{\ding{118}\xspace \textsf{#1}\xspace$\bm{\rightarrow}$\xspace(#2):\xspace}

\def\first{({i})\xspace}
\def\second{({ii})\xspace}
\def\third{({iii})\xspace}

\definecolor{verylightgray}{gray}{0.8}

\newcolumntype{L}{l<{\hspace{1cm}}}
\newcolumntype{C}{c<{\hspace{1cm}}}
\newcolumntype{D}{c<{\hspace{0.3cm}}}

\newcommand{\cmark}{\ding{51}}%
\newcommand{\xmark}{\ding{55}}

\DeclareRobustCommand\pie[1]{
\tikz[every node/.style={inner sep=0,outer sep=0, scale=1.5}]{
\node[minimum size=1.5ex] at (0,-1.5ex) {};
 \draw[fill=white] (0,-1.5ex) circle (0.75ex); \draw[fill=black] (0.75ex,-1.5ex) arc (0:#1:0.75ex);
}
}
\def\L{\pie{0}} 
\def\M{\pie{-180}} 
\def\H{\pie{360}} 

\def\cameraReady{} 

\def\BibTeX{{\rm B\kern-.05em{\sc i\kern-.025em b}\kern-.08em
		T\kern-.1667em\lower.7ex\hbox{E}\kern-.125emX}}

\begin{document}

\date{}

\title{\Large \bf \sysname: Privacy-preserving, Asynchronous Single Sign-On}

\ifdefined\cameraReady
\author{
	{\rm Zhiyi Zhang}\\
	UCLA
	\and
	{\rm Micha\l{} Kr\'{o}l}\\
	UCLouvain
	\and
	{\rm Alberto Sonnino}\\
	Facebook Calibra \\ University College London
	\and
	{\rm Lixia Zhang}\\
	UCLA
	\and
	{\rm Etienne Rivi\`ere}\\
	UCLouvain
} 
\else
\author{}
\fi

\maketitle


\begin{abstract}
\sysname is a privacy-preserving Single Sign-On system.
It implements anonymous credentials, enables selective attribute disclosure, and allows users to prove properties about their identity without revealing it in the clear.
\sysname offers the necessary tradeoff between privacy and user accountability by allowing the recovery of a misbehaving user's identity through a strict process involving multiple authorities.
\sysname is the first single-sign-on system to combine the security of anonymous credentials with the simplicity of use of Open ID Connect.
%
It is deployed as a
WebAssembly client module that can be cached by users' browsers.
It does not require additional software or hardware and streamlines traditionally difficult tasks in anonymous credentials that are multi-device support, device theft recovery, and privacy-preserving two-factor authentication.
Our implementation using PS Signatures and WebAssembly achieves 39x to 180x lower computational cost than previous anonymous credentials schemes, and similar or lower sign-on latency than Open ID Connect.
\end{abstract}


\vspace{-2mm}
\section{Introduction}
\label{sec:introduction}


Single Sign-On (SSO) is an answer to the complexity and fragility of using individual passwords on the web, \ie, leading to reuse and leaks~\cite{ives2004domino}.
SSO enables the use of a unique identity provided by an Identity Provider (IdP). 
Users authenticate themselves to services (called Relying Parties--RP) with tokens provided by their IdP.
SSO improves overall web security~\cite{goode2012importance} and enables the generalization of good security practices such as the use of 2-factor authentication (2FA)~\cite{ometov2018multi}.

\begin{figure}[t!]
	\centering
	\includegraphics[scale=0.45]{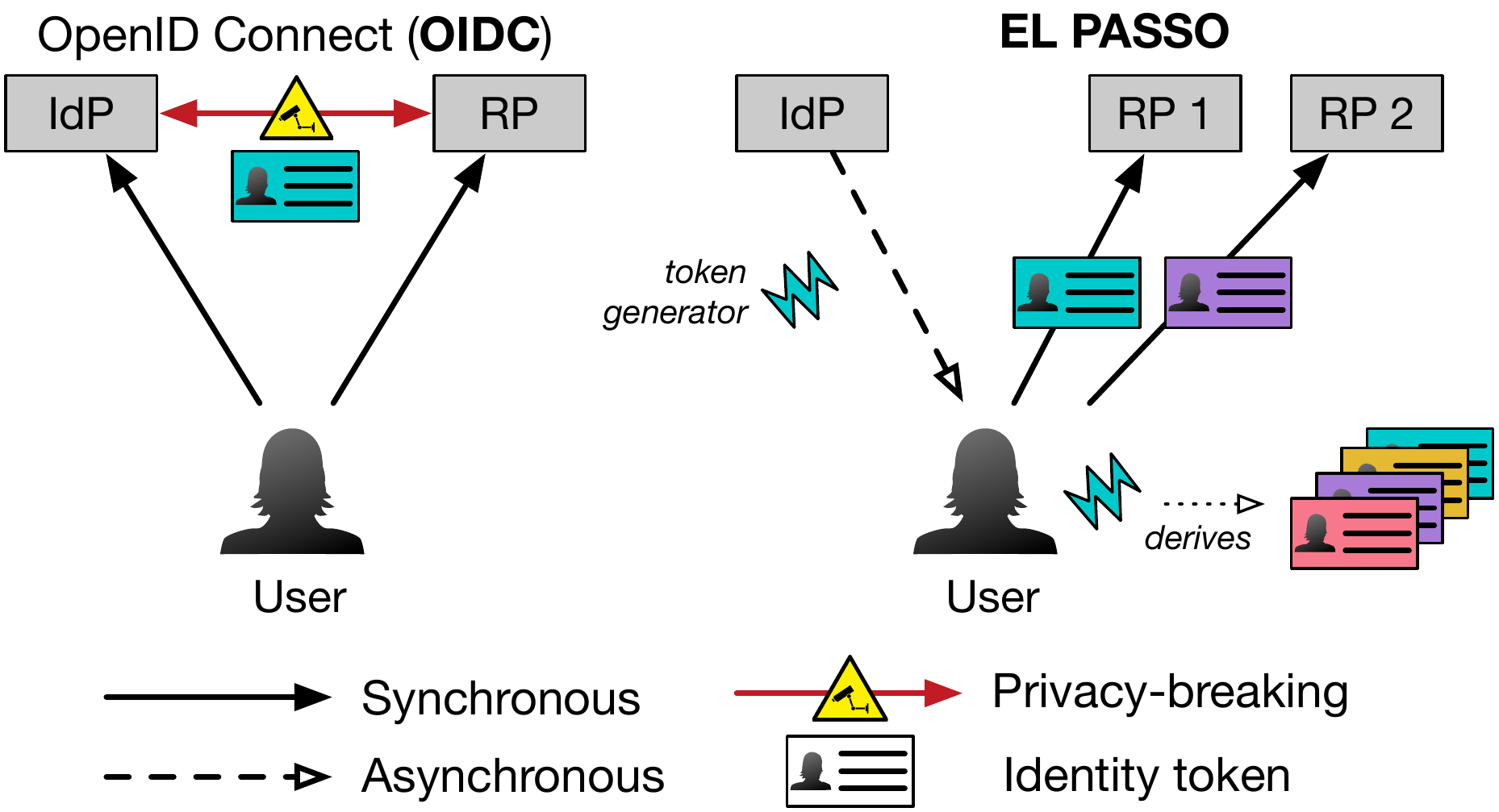}
	\caption{Sign-on in OIDC and in \sysname.}
	\label{fig:introduction_schema}
\end{figure}

\para{Limitations of OpenID Connect}
OIDC is a dominant SSO solution used by over a million websites in 2020~\cite{openid-adoption}.
Major web players such as Google or Facebook play the role of IdPs, offering so-called \emph{social login} features to RPs previously registered with their services.
However, while facilitating identity management, the wide adoption of OIDC raises concerns on users' \emph{privacy}~\cite{openid-problems,social-login-or-not-login}.
These concerns are direct consequences of the coupled mode of operation of OIDC, illustrated in \Cref{fig:introduction_schema}.
Each login request to an RP requires first an interaction between the user and the IdP for authentication and then another interaction between the RP and the IdP to validate credentials.
An IdP is, therefore, aware of its users’ every sign-on attempt and the nature of visited websites.
Similarly, any RP may learn users’ identity information (\eg, email or social media account) when interacting with the IdP, even though this is not strictly \emph{necessary} for authentication.
OIDC extensions have been proposed to increase users' privacy but only partially address these problems, as they either leak users' global identifiers to the RPs~\cite{fett2015spresso} or do not protect against the IdP~\cite{sia}.
In addition to privacy concerns, the synchronicity in OIDC impacts \emph{availability}: Users cannot connect to an RP if their IdP is offline.
This requirement of availability can prevent small organizations (\eg, digital rights NGOs) from offering an alternative to tech giants' IdPs and counter Internet consolidation~\cite{arkko2019considerations}.

\para{State of the art}
Anonymous credentials~\cite{idemix,cl,paquin2011uc,paquin2011u} have been identified as a sound basis for preserving users' privacy in SSO~\cite{abcimplem}.
They allow decoupling the interactions between the RP and the IdP and enable unlinkable authentication across RPs.
This prevents the inference by the IdP of its users' visited websites.
Unfortunately, existing authentication schemes using anonymous credentials~\cite{alpar2017irma, ursa, rannenberg2015attribute, camenisch2002design} present limitations that prevent their adoption as a drop-in replacement of OIDC.
First, they suffer from poor performance~\cite{idemix,alpar2017irma} or overheads increasing with the number of unlinkable uses and therefore with the number of RPs~\cite{paquin2011u, paquin2011uc}.
In addition, they require pre-installed specific software and manual management of the cryptographic material at the user side~\cite{ursa,idemixgen,irmamobile}. 
The tasks are too complicated for most Web users~\cite{ruoti2013confused,whitten1999johnny}, hindering the deployment of those systems at a large scale.
Finally, they do not consider multi-device scenarios, are vulnerable in case of device theft, or do not support 2FA, \ie the possibility for an RP to require a joint sign-on operation by the same user but from two different devices.

We position that, in order to be adopted widely, secure privacy-preserving SSOs must offer the same level of usability, ease of deployment, and performance as OIDC.

\para{Contributions}
We present the design, security analysis, and evaluation of \sysname, a practical privacy-preserving SSO system, illustrated in \Cref{fig:introduction_schema}.

\sysname is asynchronous and offers unlinkable authentication and the strong privacy guarantees of anonymous credentials.
The generation of authentication material by the IdP is decoupled \emph{in time} from its use by the client to sign on at some RP.
It enables minimal disclosure of information: 
	Users may share only elements necessary for a specific RP or
	provide authenticatable personal properties to an RP, such as being above a minimum age or coming from a certain geographical area, without sharing their exact age or location.

The design of \sysname acknowledges the practical consideration that \emph{unbreakable} anonymity is not desirable for many online services.
\sysname offers guardrails to the risk of digital impunity associated with minimal disclosure of information, by providing \emph{accountability} guarantees to RPs about users signing on their services.
Users convicted of fraudulent behaviors (\eg authors of hate speech or harassment in an online forum, or publishers of illegal content) can be eventually identified.
This identification obeys a strict cooperation process involving \emph{several} authorities, whose number and identity must be announced by RPs using the feature.

At the same time, \sysname aims for ease of deployment by users, RPs, and IdPs.
User-side operations are implemented as a client module in WebAssembly~\cite{haas2017bringing} received from the IdP and cached along with authentication material.
As a result, our platform does not require prior software installation or specific hardware and automatically manages cryptographic material and client code using browser built-in features. 
\sysname enables multi-device deployments: It is robust against the theft or loss of a device and the secrets it contains, and naturally supports 2FA without disclosing the user's phone or email address.

\sysname is built using PS signatures~\cite{pointcheval} and designed to limit the amount of heavy cryptographic operations required for all parties.
Our evaluation using representative user devices and RP and IdP services hosted on Amazon EC2 indicates that \sysname performance, costs, and scalability make it amenable for large-scale deployments.
Sign-on operations only require one round-trip between the user-side client and the RP, and while more computations are required at the user side than for OIDC, their CPU cost is a factor of 39x to 180x lower than for those of IRMA~\cite{idemix,alpar2017irma}, a previous platform using anonymous credentials.
This results in comparable or even lower sign-on latency compared to OIDC, \eg, only 250~ms on a laptop and 800~ms on a Raspberry~Pi representative of a mobile device.
Finally, implementations of the RP and of the IdP scale vertically and horizontally in the cloud, and allow throughput of more than 260 setup phases or more than 170 sign-on phases per second using only a 4-core VM.

\para{Outline}
We first refine our model and design goals in \Cref{sec:goals}.
We provide an overview of the design of \sysname in \Cref{sec:overview}.
We present its detailed construction, starting with background on anonymous credentials and zero-knowledge proofs in \Cref{sec:background}, followed by the protocol in \Cref{sec:construction} and its implementation in \Cref{sec:implementation}.
We provide a security analysis in \Cref{sec:sec_analysis}.
Our evaluation is given in \Cref{sec:evaluation}.
We review related work in \Cref{sec:related} and conclude in \Cref{sec:conclusion}.

\vspace{-2mm}
\section{Design Goals}
\label{sec:goals}

We start by defining our system and adversary models.
We follow up by specifying target properties for authentication, privacy, accountability, availability, and ease of deployment.

\subsection{System and Adversary Model}
\label{sec:adversaries}

Our system model aligns with that of OIDC, with three actors.
Relying Parties (RPs) are interested in allowing users to sign up with their services without creating specific accounts.
Users trust IdPs for safeguarding their identity and associated \emph{attributes}, and for providing the client code implementing user-side operations.
RPs can choose which Identity Providers (IdP) they trust for certifying the authenticity of users.
We assume that users employ a modern web browser supporting sandboxed code execution and an integrated password manager (\ie, the ability to safeguard passwords or other secrets under the user's local credentials).


We consider the following adversarial model.
IdPs are considered honest-but-curious: They do not modify the protocol or deny service.
Both IdPs and RPs may wish to break privacy guarantees or obtain authentication information allowing to impersonate users at correct RPs.
RPs may arbitrarily deviate from the protocol in addition to observing interactions with their users.
In particular, they can provide arbitrary code to run in their users' browsers.
We consider, however, that this code runs in isolation from the rest of the system, and notably from the \sysname client that is obtained from the IdP.
Users may, finally, actively attempt to abuse or bypass authentication or accountability mechanisms.
Note that the adversary can control multiple corrupted entities simultaneously, \eg set up several RPs, or a combination of users and RPs.

\begin{table}[]
\resizebox{\linewidth}{!}{%
\footnotesize

\newcolumntype{C}{>{\raggedright\let\newline\\\arraybackslash\hspace{0pt}}m{0.4\linewidth} }
\newcolumntype{D}{>{\raggedright\arraybackslash} m{0.625\linewidth} }

\begin{tabular}{CD}
\toprule
\multicolumn{2}{l}{\textbf{Authentication}}                        \\
\midrule
personal authentication        & only legitimate IdPs users can authenticate \\
intra-RP linkability           & prevent creation of Sybils within a domain\\
\midrule
\multicolumn{2}{l}{\textbf{Privacy}}                \\
\midrule
selective attributes disclosure & user only discloses necessary attributes \\
provable personal properties   & user attributes' properties attested by IdP \\
tracking protection            & IdP not aware of user's sign-ons \\
inter-RP unlinkability         & sign-ons across multiple RPs cannot be linked \\
\midrule
\multicolumn{2}{l}{\textbf{Accountability}}                      \\
\midrule
reliable identity retrieval    & misbehaving users identity can be revealed \\
\midrule
\multicolumn{2}{l}{\textbf{Deployment}}                          \\
\midrule
asynchronous authentication & can sign on even if IdP temporarily unavailable \\
no RP registration & RP does not have to register with IdP \\
browser-only & no software pre-installation required \\
multi-device support & support device theft \& two-factor authentication \\
\bottomrule
\end{tabular}
}
\vspace{-1mm}
\caption{Target properties of \sysname.
\vspace{-2mm}}
\label{tab:sso_properties}
\end{table}

\subsection{Target Properties for \sysname} \label{sec:properties}


\noindent
\sysname provides the properties listed in \Cref{tab:sso_properties}:

\para{Authentication}
\sysname only allows legitimate users registered with an IdP to sign up and on with an RP.
It prevents any other entity in the system from impersonating existing user accounts created at RPs\footnote{This includes IdPs, who are not allowed to possess material for signing on as a specific user at an RP, to prevent account abuse in case of a data leak.} (\emph{personal authentication}).

Authentication requirements also include the prevention of Sybil identities, disallowing a user from creating multiple identities for the same domain.
RPs can detect authentication attempts made with credentials issued by an IdP for the same user (\emph{intra-RP linkability}).

\para{Privacy}
\sysname targets \emph{minimal disclosure of information}, \ie the ability for users to control the amount of information about their profile they wish to share with RPs.
A user can select which of their attributes (\eg email address, but not last name) should be revealed to an RP.
Note that a user still benefits from personal authentication when sharing \emph{none} of their personal attributes (\emph{selective attributes disclosure}).
A user may even decide to only share authenticatable certifications of \emph{properties} about their attributes, without disclosing their values (\emph{provable personal properties}).
For instance, the 2005 Gambling Act of the United Kingdom~\cite{gamblingact} requires users of online casinos to be at least 18 years old, and holds online services responsible to enforce the regulation.
In this example, \sysname can provide a certificate that a specific user is over 18 years old, while their actual age does not need to be revealed.

\sysname prevents the tracking of users' activity.
It is unfeasible for IdPs to \emph{track} the sign-ons activity of their users onto different RPs, to prevent profiling and the resulting leakage of personal information~\cite{krishnamurthy2007measuring} (\emph{tracking protection}).
In addition, in the absence of common information, it is impossible to correlate multiple accounts created from the same credential on different RPs (\emph{inter-RP unlinkability}).
For instance, an account on one RP disclosing the real name of a user cannot be correlated with another account, for the same user but at another RP, that only revealed the user's address.

\para{Accountability}
\sysname enables \emph{accountability} of users, mitigating the risks associated with anonymous identities, and enabling privacy preservation for services such as online democracy.
If a user engages in reprehensible behavior such as publishing illegal content or harassment, a set of authorities can eventually collaborate and hold them accountable, in cooperation with the IdP (\emph{reliable identity retrieval}).
RPs must announce the use of accountability, the set of authorities, and the threshold number of authorities strictly necessary for re-identification.
RPs can validate that their users provide the necessary identity recovery material upon sign-on.

\para{Deployment}
SSO services become a critical part of many information systems~\cite{sun2011makes}.
Even large, highly redundant systems may experience downtimes, as exemplified by the recent 14-hour disruption of Facebook's services in March 2019~\cite{facebook_outage} or the Amazon AWS outage in 2018~\cite{awsdown}.
In \sysname, a user does not need to be authenticated by the IdP \emph{each time} they sign on with an RP; instead, users acquire their credentials periodically and can connect to RPs even when the IdP is temporarily offline (\emph{asynchronous authentication}).

RPs do not need to register with IdPs to be able to trust authentication information, and it is impossible for IdPs to impersonate each other.
The sign-on process is universal: RPs do not need specialized operations for a specific IdP (\emph{no RP registration}).
This improves system automation and mitigates Internet consolidation~\cite{arkko2019considerations}, as RPs becomes more independent and new, smaller IdPs can enter the market more easily.


On the user side, \sysname does \emph{not} require specific hardware (\eg, a trusted execution environment), physical device (\eg, an external fingerprint reader or a smart card) or extra network services to offer its functionalities.
It does not require, either, the installation of a specific software client, and all user-side code runs as sandboxed code inside their web browser (\emph{browser-only}).

Finally, \sysname supports multi-device scenarios.
It enables users to easily register new devices (\eg, laptop, phone, tablet) and supports easy identity recovery in case of the theft of one device.
It natively supports 2FA: An RP may request and assess that users connect from two different devices in order to sign on their services (\emph{multi-device support}).


\vspace{-3mm}
\section{Overview of \sysname}
\label{sec:overview}

\begin{figure}[t]
\centering
\includegraphics[scale=0.45]{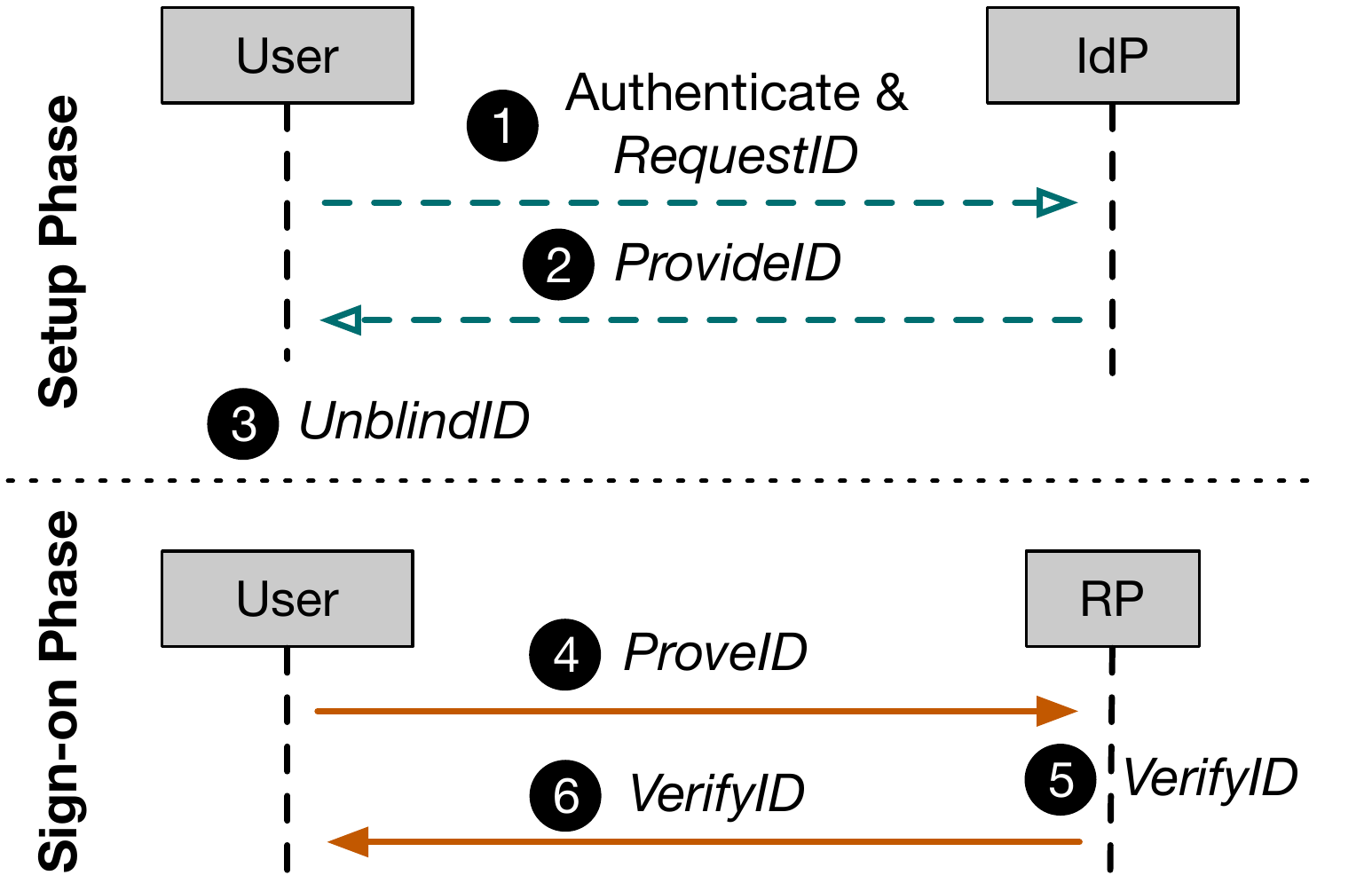}
\vspace{-1mm}
\caption{\sysname uses two phases; (i) a \emph{setup phase} where the user obtains a credential from their identity provider, and (ii)~a \emph{sign-on phase} where the user proves possession of their credential to anonymously authenticate to websites.
\vspace{-2mm}}
\label{fig:overview}

\end{figure}

A fundamental design principle of \sysname is the avoidance of synchronous communication between RPs and IdPs.
User-side clients derive, instead, RP-specific tokens based on material previously obtained from the IdP (\Cref{fig:introduction_schema}).
The generation and use of tokens are divided into two asynchronous phases (\Cref{fig:overview}).
In the \emph{setup phase}, the client obtains an anonymous credential from the user's IdP.
In the \emph{sign-on phase}, the client prepares an RP-specific derivation of this credential based on  what information the user decides to disclose, and proves the authenticity of this client to the RP.
The setup phase is executed periodically (\eg once every few days), while the sign-on phase is executed each time the user logs in or creates an account on an RP.
%

\para{Setup phase}
The user first authenticates to their IdP and runs \textsf{RequestID} to request a credential from the IdP~(\ding{202}) over a random attribute $s$ acting as user secret; $s$ is hidden from the IdP.
Users can also specify which information $\mathit{info}$ they need to be embedded in their credentials, such as their email address, names, or age.
If the IdP successfully authenticates the user, verifies the user's knowledge of the secret $s$, and knows the requested information about the user, it runs \textsf{ProvideID} to issue a credential embedding that information as well as a long-term pseudonym $\gamma$ unique to the user and a timestamp $\mathit{tp}$ marking the expiration date of the credential~(\ding{203}).
The IdP also keeps $h^\gamma$ for the user in their database, where $h$ is a public parameter.
The user-side client locally runs \textsf{UnblindID} to unblind the received credential~(\ding{204}).
Importantly, the IdP (or anyone else) is unable to use the credential on behalf of the user as this would require knowing the user secret $s$.

\para{Sign-on phase}
The user-side client connects to the RP and executes \textsf{ProveID} to prove knowledge of the credential issued by the IdP (\ding{205}).
In this process,
\first~the client locally randomizes the credentials so that even if an attacker observes both the RP and the IdP it cannot link the credentials to a specific user of this IdP.
\second~The client provides the RP with the randomized credential and the expiration time $\mathit{tp}$, and selectively discloses any subset of information $\mathit{info'}$ embedded in the credential, enabling \emph{selective attributes disclosure}.
The client may also generate and include proofs about \emph{properties} of the attributes they do not wish to share in the clear, enabling \emph{provable personal properties}.
\third~The client locally generates a group element $\zeta$ uniquely derived from the user's secret $s$ and the RP's DNS domain name, and proves in zero-knowledge its correctness; the RP uses this group element as the device-specific user ID.
Once the RP verifies the credential along with the proofs, it considers the user as authenticated~(\ding{207}).

To verify that the credential was initially issued by a trusted IdP, an RP must collect the public key of this IdP.
Since the setup and sign-on phases are asynchronous, the RP can fetch the public key from the IdP domain directly, by issuing a GET request over \texttt{https} to the IdP\footnote{The IdP cannot correlate a previous setup phase with the current sign-on phase at this RP. This contrasts with synchronous designs such as SPRESSO~\cite{fett2015spresso} where the collection of public parameters must happen via Tor to prevent time-based attacks by the IdP.}.
This requires the IdP to be online for a sign-up operation (creation of a new account), but the RP can cache this public key for future sign-ons, enabling \emph{asynchronous authentication}.
Since the credentials are randomized (as part of \textsf{ProveID}) and the alias $\zeta$ changes for each RP without leaking $s$, a user can employ the same credential to authenticate to different RPs.
At the same time, different sign-ons to different RPs are unlinkable even if observed by the same adversary.
A user cannot create multiple accounts with a single RP since $\zeta$ is bound to the user's secret embedded in a credential and the RP's DNS domain name; it is infeasible to create two different $\zeta$ over the same RP's DNS domain name from a single credential.

To mitigate risks of correlation of requests at different RPs by the adversary, the timestamp $\mathit{tp}$ should be rounded or limited to denominations fixed by the IdP\footnote{Alternatively, instead of revealing the time-stamp upon execution of \textsf{ProveID}, the client could prove in zero-knowledge that the timestamp is greater than the current date (but this requires a potentially expensive range-proof).}; \eg specifying an expiration day, but omitting more detailed information such as hours, minutes and seconds.

Finally, if support for \emph{reliable identity retrieval} is required by the RP, the client must provide and prove the correctness of an \elgamal encryption $E$ of their long-term pseudonym $\gamma$ encrypted under the public key of specific decryption authorities, as part of the \textsf{ProveID} operation.
If the user misbehaves, the RP discloses $E$ to these decryption authorities, which decrypt it to obtain $h^\gamma$, and then collaborate with the IdP to recover the identity of the user.
\sysname supports flexible key management for decryption authorities---the ciphertext is typically encrypted using \emph{threshold encryption}, where at least a threshold number of decryption authorities are needed to recover $h^\gamma$ and, therefore, the user's identifier.

\para{Multi-device support}
A user may add a new device and use it to connect to RP accounts created with any of their older devices.
The new device needs to receive the secret $s$ without leaking it to any third party; this can be achieved as follows.
The new device generates an ephemeral public/private key pair and sends the public key to the IdP.
The user confirms the new device at the IdP using one of the older devices, and the user-side client encrypts $s$ under the new device's public key.
To ensure the integrity of the public key, the user inputs a number on both devices used as salt.
The IdP sends the encrypted secret to the new device allowing it to request credentials over $s$.

\para{Two-factor authentication}
\sysname allows RPs to require two-factor authentication (2FA), \ie that users connect from two different devices for attesting their authenticity.
Under the principle of minimal disclosure of information, 2FA does not require revealing an email address or phone number, but only to use two different previously-enabled devices.
This requires, in addition to secret $s$, a device-specific secret $s_d$.
The client includes $s_d$ during the setup phase making it a part of the credentials.
When the user connects with a given device to the RP for the first time,
they provide $\zeta$ and generate an RP- and device-specific pseudonym $\zeta_d$ derived from $s_d$ and the RP's DNS domain name.
Similarly to $\zeta$, $\zeta_d$ are unlinkable across domains and cannot be re-used by a malicious RP.
The RP is able to link the new device to the user account using $\zeta$ and adds $\zeta_d$ to the list of authorized devices.
Subsequent logins using the same device requires only providing $\zeta_d$ and do not involve additional overhead.
When requiring 2FA, an RP simply checks that two subsequent logins are performed from devices with different values of $s_d$.

%

\para{Device theft recovery}
A user can declare the loss of a device to their IdP.
The IdP will stop issuing credentials for that device.
A thief able to unlock the secret storage of the stolen device's browsers would be able to connect to RPs, unless 2FA is required, but only until the IdP credential expires.
It will not be able to authorize new devices.
Users do not lose access to their RP accounts as long as they hold at least one device (or two devices, if 2FA is required).
A user can replace their secret $s$ using the following procedure.
The user contacts the IdP and asks for credentials on a new, blinded $s'$; from now on, the IdP will not renew credentials for $s$ to preserve sybil resistance.
The client connects to the RP and presents: credentials over the old, expired $s$, $\zeta(s)$, credentials over the new $s'$ and $\zeta(s')$.
The RP replaces $\zeta(s)$ by $\zeta(s')$, and stops accepting credentials on $s$.


\vspace{-2mm}
\section{Building Blocks}
\label{sec:background}

We present our building blocks, anonymous credentials and zero-knowledge proofs, and our cryptographic assumptions.

\vspace{-1mm}
\subsection{Anonymous Credentials}

Anonymous credentials~\cite{amac, pointcheval} allow the issuance of credentials to users, and the subsequent unlinkable revelation to a verifier.
Users can selectively disclose some of the attributes embedded in the credential or specific functions of these attributes.
\sysname requires a credential scheme providing short and computationally efficient credentials, re-randomization, unlinkable multi-show selective disclosure, and blind issuance~\cite{amac}.
An anonymous credential scheme can be defined by the set of algorithms below.
\begin{description}[leftmargin=1em, labelindent=0em]
\item[\newdef{Cred.Setup($1^\lambda$)}{$\mathit{pp}$}] define the system parameters $\mathit{pp}$ with respect to the security parameter $\lambda$. These parameters are publicly available.

\item[\newdef{Cred.KeyGen($\mathit{pp}$)}{$\mathit{sk}, \mathit{pk}$}] run by the authority to generate their own secret key $\mathit{sk}$ and public key $\mathit{pk}$ from the public parameters $\mathit{pp}$.

\item[\newdef{Cred.Issue($sk, M_h,M_p,\phi$)}{$\sigma$}] interactive protocol between the user-side client and the authority; the client obtains a credential $\sigma$ embedding the set of public attributes $M_p$ and the set of hidden attributes $M_h$ if they satisfy the statement~$\phi$. \textsf{Cred.Issue} is composed of three algorithms:

\begin{description}
\item\newdef{Cred.PrepareBlindSign($pk, M_h, \phi$)}{$d, \Lambda, \phi$} run by the client to generate the blind factor $d$, and the cryptographic material $\Lambda$ (embedding $M_h$) over which the authority blindly issues a credential.

\item\newdef{Cred.Sign($sk, M_p, \Lambda,\phi$)}{$\tilde{\sigma}$} run by the authority to issue the blinded credentials $\tilde{\sigma}$ over $M_p$ and $\Lambda$, using their private key $\mathit{sk}$.

\item\newdef{Cred.Unblind($d, \tilde{\sigma}$)}{$\sigma$} run by the client to unblind $\tilde{\sigma}$ (using the factor $d$) to retrieve the credential $\sigma$.
\end{description}

\item[\newdef{Cred.Prove($pk, M_p, M_h, \sigma, \phi'$)}{$M_p, \Theta, \phi'$}] run by the client to compute a proof $\Theta$ proving possession of a credential $\sigma$ certifying that the private attributes $M_h$ and the public attributes $M_p$ satisfy the statement $\phi'$\footnote{Note that $\phi'$ may be different from $\phi$.}.

\item[\newdef{Cred.Verify($\mathit{pk}, M_p, \Theta, \phi'$)}{$b$}] run by any third party verifier to verify that the credential represented by the cryptographic material $\Theta$ embeds $M_p$ as well as hidden attributes satisfying the statement $\phi'$, using the public key $\mathit{pk}$ of the issuing authority.
\end{description}
All algorithms receive the security parameter $\lambda$ as an input but we show it explicitly only for \textsf{Cred.Setup}.
\sysname uses \ps~\cite{pointcheval} as the underlying credentials scheme as it uses short, and computationally efficient credentials.
We use \ps for the generation of credentials by the IdP, and for the verification of credentials by RPs on both known messages (\eg, timestamp $\mathit{tp}$) and hidden messages (\eg, user's secret $s$).

\vspace{-1mm}
\subsection{Zero-knowledge Proofs}

Zero-knowledge proofs are protocols allowing a \emph{prover} to convince a \emph{verifier} that it knows a secret value $x$, without revealing any information about that value.
The prover can also convince the verifier that they know a secret value $x$ satisfying some statements $\phi$.
Anonymous credentials extensively employ zero-knowledge proofs to provide users with certified secret values; users are successively able to prove to third party verifiers that they hold secret values certified by specific credentials issuers, and prove statements about those values without disclosing them.
This enables, for instance, the property of \emph{provable personal properties}.
A credential issuer may provide a user with a secret value $x=20$ representing their age; the user can then prove in zero-knowledge to a verifier that a specific credential issuer certified that their age is larger than 18, without revealing their real age $x$.

\sysname uses non-interactive zero-knowledge proofs (NIZK) to assert knowledge and relations over discrete logarithm values.
These proofs can be efficiently implemented without trusted setups using sigma protocols~\cite{schnorr}, which can be made non-interactive using the Fiat-Shamir heuristic~\cite{RFC8235} in the random oracle model.

\vspace{-1mm}
\subsection{Cryptographic assumptions}
\label{sec:crypto-assumptions}

\sysname inherits the same cryptographic assumptions as \ps, which requires groups $(\mathbb{G}_1,\mathbb{G}_2,\mathbb{G}_T)$ of prime order $p$ with a bilinear map $e:\mathbb{G}_1 \times \mathbb{G}_2 \rightarrow \mathbb{G}_T$ and satisfying \first \emph{Bilinearity}, \second \emph{Non-degeneracy}, and \third \emph{Efficiency}.
We use type-3 pairings because of their efficiency~\cite{galbraith2008pairings}, and therefore rely on the XDH assumption which implies the difficulty of the Computational co-Diffie-Hellman (co-CDH) problem in $\mathbb{G}_1$ and $\mathbb{G}_2$, and the difficulty of the Decisional Diffie-Hellman (DDH) problem in $\mathbb{G}_1$~\cite{bls}.
We also rely on a cryptographically secure hash function $\hashtopoint$, hashing a string into an element of $\mathbb{G}_1$; \ie applying a full-domain hash function to hash strings into elements of $\mathbb{G}_1$ (such as BLS~\cite{bls}).


\vspace{-2mm}
\section{\sysname Construction}
\label{sec:construction}


We present the construction of \sysname satisfying all properties described in \Cref{sec:properties}, and then discuss how to simplify it when \emph{reliable identity retrieval} is not required or if the user wishes to sign on as guest without establishing an identity with the RP, and how to support login with multiple devices.
We discuss the implementation of the protocol steps in \Cref{sec:implementation} and their security guarantees in \Cref{sec:sec_analysis}.
\sysname primitives (see \Cref{fig:overview}) are defined as follows:

\para{Bootstrapping the IdP} The following algorithms are executed only once, when bootstrapping the IdP.
\begin{description}[leftmargin=1em, labelindent=0em]
\setlength\itemsep{.3em}
\item[\newdef{Setup($1^\lambda$)}{$\mathit{pp}$}] output \textsf{Cred.Setup($1^\lambda$)}. \\
$\triangleright$ Describe the publicly-available system parameters with respect to the security parameter $\lambda$.

\item[\newdef{KeyGen($\mathit{pp}$)}{$\mathit{sk}, \mathit{pk}$}] output \textsf{Cred.KeyGen($\mathit{pp}$)}. \\
$\triangleright$ Run by the IdP to generate their own secret key $\mathit{sk}$ and public key $\mathit{pk}$ from the public parameters $\mathit{pp}$.
\end{description}

\para{Setup phase} We describe the algorithms implementing the setup phase of \sysname; these algorithms are executed periodically, when the user requests a credential from the IdP.
\begin{description}[leftmargin=1em, labelindent=0em]
\item[\newdef{RequestID($s$)}{$\Lambda$}] set $M_h=s$ and $\phi=\mathit{true}$; run $(d, \Lambda, \perp)$ = \textsf{Cred.PrepareBlindSign($M_h, \phi$)};  output $\Lambda$. \\
$\triangleright$ Run by the user-side client to request a credential from the IdP, generating the cryptographic material $\Lambda$ embedding the user secret $s$ along with the proof.
The blinding factor $d$ will be kept by the client for later use.

\item[\newdef{ProvideID($\mathit{sk}, \gamma, \mathit{info}, \mathit{tp}, \Lambda$)}{$\tilde{\sigma}$}] set $M_p=(\gamma, \mathit{tp}, \mathit{info})$; output $\tilde{\sigma}$ = \textsf{Cred.Sign($\mathit{sk}, M_p, \Lambda, \mathit{true}$)}. \\
$\triangleright$ Run by the IdP to provide the client with a blinded credential $\tilde{\sigma}$ over $\Lambda$, the user identifier $\gamma$, and some user attribute $\mathit{info}$; the credential has an expiration date $\mathit{tp}$, and is produced from the IdP's secret key $\mathit{sk}$.

\item[\newdef{UnblindID($d, \tilde{\sigma}$)}{$\sigma$}] output $\sigma = \textsf{Cred.Unblind($d, \tilde{\sigma}$)}$. \\
$\triangleright$ The client locally unblinds the credential $\tilde{\sigma}$ using the blinding factor $d$, and outputs the credential $\sigma$.
\end{description}

\para{Sign-on phase} We describe the algorithms implementing the sign-on phase; these algorithms are executed each time the users logs in or creates an account on an RP.

\begin{description}[leftmargin=1em, labelindent=0em]
\item[\newdef{ProveID($\mathit{pk}, \sigma, \gamma, \mathit{info}, \mathit{tp}, \mathit{domain}, y$)}{$\Theta, \phi'$}] split $\mathit{info}$ into $\mathit{info_p}$ and $\mathit{info_h}$, respectively containing the attributes to disclose and to hide from the RP.
Set $M_p=(\mathit{info_p}, \mathit{tp})$ and $M_h = (s, \gamma, \mathit{info_h})$;
pick a random $\epsilon \leftarrow \mathbb{F}$ and compute the \elgamal ciphertext $E = (g^\epsilon, y^\epsilon h^\gamma)$, where $g$ and $h$ are generators of $\mathbb{G}_1$, and $y$ is the aggregated public key of the decryption authorities;
compute
$$\tilde{h} \leftarrow \hashtopoint(\mathit{domain}) \; ; \; \zeta \leftarrow \tilde{h}^s$$
(where $\hashtopoint$  is a hash function as defined in \Cref{sec:crypto-assumptions})
and the statement
$$\phi' \leftarrow  \{ E = (g^\epsilon, y^\epsilon h^\gamma) \; \land \; \zeta=\tilde{h}^s  \; \land \;  f(\mathit{info_h})=1\}$$
compute $(\Theta, M_p, \phi')$ = \textsf{Cred.Prove($\mathit{pk}, M_p, M_h, \sigma, \phi'$)}; output $(\zeta, \Theta, M_p, \phi', f)$\\
$\triangleright$ Run by the client to show the RP a proof of correctness of user ID $\zeta$ and identity retrieval token $E$, and the ownership $\Theta$ of a credential $\sigma$ whose attributes satisfy the statement $\phi'$; this proof is generated from the RP's public domain $\mathit{domain}$, and from the parameters $(\mathit{pk}, \gamma, \mathit{tp}, y)$.
The subset of hidden attributes $\mathit{info_h}$ satisfy the function $f$.

\item[\newdef{VerifyID($\mathit{pk}, M_p, \Theta, \phi', \mathit{domain}, y$)}{$b$}] compute $\tilde{h}=\hashtopoint(\mathit{domain})$ and use it to execute \textsf{Cred.Verify($\mathit{pk}, \Theta, \phi'$)}; output $b=1$ if \first the verification passes, \second the time-stamp $\mathit{tp}$ is not expired, and \third the $\zeta$ and \elgamal ciphertext $E'$ are correctly formed; otherwise output $b=0$. \\
$\triangleright$ Run by the RP to verify that $\Theta$ is a proof of knowledge of a valid credential (issued by the IdP identified by the public key $\mathit{pk}$) whose attributes satisfy the statement $\phi'$, and user ID $\zeta$ and identity retrieve token $E$ are correct; the proof is verified using $(M_p, \mathit{domain}, y)$.
\end{description}

\para{Removing reliable identity retrieval}
In case support for reliable identity retrieval is not required by the RP (see \Cref{sec:overview}), we can simplify the sign-on phase of the above scheme by omitting the ciphertext $\varepsilon$; the statement $\phi'$ would then become $\phi'=\{ \zeta=h^s  \; \land \;  f(\mathit{info_h})=1\}$, and the zero-knowledge proof $\Theta$ shorter by two field elements (if implemented, for instance, using Schnorr’s protocol~\cite{schnorr}).

\para{Login as guest}
In case the user wishes to sign on as guest without establishing a permanent user identifier with the RP, and if the RP allows such guest sign-ons, we can simplify our scheme by omitting the group element $\zeta$; the statement $\phi'$ would then become $\phi'=\{ E = (g^\epsilon, y^\epsilon h^\gamma) \; \land \;  f(\mathit{info_h})=1\}$, which shortens the proof $\Theta$ by one group element. As a result, the RP has no way to distinguish multiple sign-ons from the same user (this follows directly from the unlinkability properties of the underlying credentials scheme).
The interaction between the user and the RP is still anonymous and accountable.

\vspace{-2mm}
\section{Implementation}
\label{sec:implementation}
\vspace{-1mm}

The RP, IdP, and user-side client are implemented in C++ using the MCL library~\cite{mcl} and Google's Protocol Buffer~\cite{protobuf}.
The C++ implementation of the client is ported to JavaScript code using WebAssembly (Wasm)~\cite{wasm}\footnote{WebAssembly (Wasm) is an open standard that defines a portable binary-code format for executable programs, a corresponding textual assembly language, as well as interfaces for facilitating interactions between these programs and their host environment. Wasm code can run natively in all major web browsers. Various compilers allow transforming high-level language source code (\ie C++, Rust) into a binary file which runs in the same sandbox as regular JavaScript code.}.
The use of C++ and Wasm allows our implementation to provide both high efficiency and the ability to be delivered as a web resource.
The footprint of executables is 178~KB for the RP, 237~KB for the IdP, and 264~KB for the client, including  the Wasm binary and JavaScript `glue' code.
All user-side operations (\ie, cryptographic operations, secret sharing) are automatically handled by the Wasm client running in the browser, resulting in a deployment complexity that is similar to that of OIDC.

\para{Obtaining credentials and client code}
IdP operations (authentication and \textsf{RequestID}) are accessed through a web page hosted in the IdP domain.
Users connect to this page to obtain credentials, and the content of the page is then cached locally, including the javascript code and the Wasm client module.
We leverage the fact that Wasm modules are fully cacheable by the browser~\cite{wasmcaching} and can be marked as immutable~\cite{cachecontrol}.
When a RP wishes to authenticate a user, it redirects to the authentication page of their IdP (selected by this user from a list of IdPs trusted by this RP).
The user is able to check that the authentication page URL is, indeed, part of the domain of their IdP.
The user is then presented with an interface to select which attributes and provable properties they wish to present to this RP.
In the majority of cases, the authentication page and Wasm module will be cached, use locally stored credentials, and there will be no direct interaction with the IdP, enabling the property of \emph{asynchronous authentication}.
When there is no cached version we favor continuity of experience for the user and allow a synchronous redirection to happen towards the page hosted by the IdP, as done in SPRESSO~\cite{fett2015spresso}.
This pragmatic approach enables continuity of experience at the cost of a minor risk on tracking protection.
Privacy-conscious users can, and should, avoid this risk by systematically pre-authenticating with the IdP. 

\para{Credential/Secret storage}
The client module needs to store and later retrieve user secrets (global $s$ and device-specific $s_d$) as well as the credentials received from the IdP.
The client runs as a sandboxed Wasm module, unable to interact with the outside world (\eg, the file system).
We leverage instead the password manager service provided by the browser to store secrets and credentials securely.
Users have to locally authenticate with their browser (\eg using a password, fingerprint, or face recognition) in order to grant the client access to this information.
User do not need to be exposed to secrets $s$ and $s_d$; They only need to know their local password and the password used at their IdP, as when using OIDC.
The password manager only accepts get requests for the same domain that stored data initially, effectively protecting against attacks that would attempt to redirect the user to a fake authentication page or bypass user's scrutiny of this page domain name (\eg, using a typosquatting attack~\cite{szurdi2014long}).


%

\para{Anonymous credentials}
We implement \sysname using \ps as the underlying credential system because of its short credentials and efficient verification.
Our prototype is implemented over the curve BLS12-381~\cite{bls12-381}.


\para{State size}
IdPs store their own key pair and a $h^\gamma$ (32 Bytes) for each of their users; RPs store the public key of each IdP they trust, aggregated public keys $y$ (32 Bytes) of decryption authorities they wish to use, as well as ciphertext $E$ (64 Bytes) and group element $\zeta$ (32 Bytes) for each of their users.
Since our implementation is based on \ps, the size of the public key of the IdP increases linearly with the number of attributes, ranging from 466 Bytes (for 3 attributes) to 2,166 Bytes (for 20 attributes).
Users store all the input parameters of \textsf{ProveID}, that is, their attributes ($s$, $\gamma$, $\mathit{info}$, $\mathit{tp}$), their credential $\sigma$ (64 Bytes), the public key $\mathit{pk}$ of the IdP who issued the credential, and the public key $y$ (32 Bytes) of each of the decryption authorities (when \emph{reliable identity retrieval} is required).
All parties are aware of the public parameters (generated by \textsf{Setup}).
In the simplest scenario where there exists one IdP, one user, and one RP, assuming there are 3 attributes and that reliable retrieval is required, the total state required at the IdP, the user, and the RP are 562 Bytes, 630 Bytes, and 594 Bytes, respectively.


\vspace{-2mm}
\section{Security Analysis}
\label{sec:sec_analysis}
\vspace{-1mm}

We start by analyzing the security of \sysname against the properties defined in \Cref{sec:properties}.
We discuss next how to extend the security and the fault model.
Finally, we discuss known attacks against incorrect OIDC implementations and the sensitivity of \sysname to similar attack vectors.

\subsection{Achievement of Design Goals}

We argue that \sysname satisfies the design goals described in \Cref{sec:properties} under the adversary described in \Cref{sec:adversaries}. 

\para{Authentication}
\sysname preserves authentication against malicious users.
Authentication relies on the unforgeability of the underlying credential system---it is unfeasible for malicious users to execute \textsf{ProveID} without holding a valid credential issued by an IdP.
Furthermore, IdPs cannot take over accounts that users created with RPs as \textsf{ProveID} requires to provide the RP with a group element $\zeta$ that is uniquely derived from the unique user secret $s$, and that is persistent across authentications.
The blind issuance and zero-knowledge properties of the underlying anonymous credentials system guarantee that the user secret attribute $s$ is not revealed to the IdP (nor to any other party); hence an IdP or RP under the control of the adversary cannot impersonate existing users and access existing accounts at RPs.
In case one of the devices is compromised and the attacker is able to access the password manager on the victim user's device, this attacker will be able to log in to RPs, but only until the credentials expire.
Similarly as for OIDC, this risk is mitigated by the use of 2FA and the ability to replace the secret, both features that \sysname enables.

\para{Privacy Protection}
The privacy guarantees of \sysname rely on the security (blind issuance and unlinkability) of the underlying credential scheme, and on the zero-knowledge property of the selected NIZK scheme.
The blind issuance property of the underlying credential scheme ensures that \textsf{RequestID} does not leak any information about the secret attribute $s$ to an honest-but-curious IdP; and zero-knowledge ensures that \textsf{ProveID} reveals to RPs no additional information about users' attributes than what is selectively disclosed by the user.
To complete the argument, note that \first revealing $\zeta$ does not leak $s$, and $\zeta$ changes indistinguishably for each website's domain (assuming a random oracle); and \second  the ciphertext $E$ hides $\gamma$ (by the security of \elgamal encryptions) under the assumptions of the underlying cryptographic primitives.

\para{Accountability}
\sysname guarantees reliable identity retrieval against misbehaving users.
\textsf{ProveID} requires users to provide RPs with a ciphertext $E$, and prove in zero-knowledge that it is correctly formed; therefore, RPs can check that $E$ is a valid encryption of the user's long-term identifier $h^\gamma$ when the user signs in, even without decrypting it.
If this user later misbehaves, the RP can report $E$ to a subset of the decryption authorities (identified by the public key $y$) that can recover the user's long term identifier $h^\gamma$, and then collaborate with the IdP to recover the user's real-world identity.
When executing \textsf{ProveID}, users can only disclose or prove statements about attributes that are certified by the IdP (\ie they cannot add, remove, or modify attributes); this follows from the unforgeability of the underlying credential system, and enables provable personal properties.



\subsection{Limitations and Stronger Adversaries}

We now discuss limitations and possible extensions to the adversary model defined in \Cref{sec:adversaries}.

\para{Actively malicious IdPs}
\sysname only considers honest-but-curious IdPs; that is, actively malicious IdPs are not part of the threat model.
An actively malicious IdP~\cite{mainka2016not} can break authentication by self-issuing credentials and create fake identities; and can break accountability by refusing to cooperate with decryption authorities.
It cannot, however, access an existing user account with an RP, as this is bound to the user secret $s$.
Furthermore, an actively malicious IdP cannot be trusted to deliver the user-side Wasm module.
We defer the protection against malicious IdPs to future work.
One possibility would be to extend distributed SSO solutions~\cite{josephson2004peer,chen2005threspassport}, where a set of IdPs must collectively authenticate a user and provide it with \emph{shares} of their identity, and to rely on trusted standardization authorities for delivering the Wasm module.
In a first iteration, this role could be played by organizations such as the W3C or digital rights NGOs.
Eventually, the inclusion of this code in Web browsers, e.g. by the Mozilla foundation in Firefox, would be a more solid approach.




\para{User device under control of the adversary}
A malicious RP may inject code at the side of a honest user, but this code is sandboxed from rest of the environment and in particular against the Wasm client module.
The RP may, for instance, set up a phishing attempt by displaying a fake authentication page to the user and using a corrupted Wasm client module.
The user may fail to notice that the URL does not match that of their IdP.
We make the assumption that the password manager of the browser is secure and does not reveal secrets and credentials, as the domain name for the storage and retrieval of this information does not match.
There have been examples, however, of successful attacks against password managers~\cite{li2014emperor,silver2014password}.
The risk for \sysname is the same as for OIDC in this case.
The solution, besides employing more secure designs for password managers~\cite{mccarney2012tapas}, is to systematically require the use of 2FA.
As for OIDC implementations, which do not require re-authenticating with the IdP using 2FA for \emph{every} sign-on operations, the frequency of 2FA is a compromise between convenience and security, and can be adjusted with timestamp $\mathit{tp}$ upon issuance of the credential.
We note, however, that the asynchronous nature of \sysname prevents re-using some of the existing safety checks performed at the IdP in OIDC, such as checking for unusual origin locations of authentication requests.
Such safety checks must, instead, be implemented at the RP side, possibly using provable personal properties.
A second solution is to use secure login solutions such as the W3C Web Authentication protocol, WebAuthn.
This standard requires the use of an external trusted device, the authenticator, for storing private keys and verifying users identity (\eg, using biometrics or passcodes).

\subsection{Sensitivity to Known Attacks on OIDC}
We discuss attacks and exploits against incorrect implementations of OIDC~\cite{fett_extensive_2019}, and the extent to which \sysname's design prevents similar attack vectors.

A first category of attacks exploits the coupling between the RP and the IdP in OIDC.
IdP Mix-Up Attacks~\cite{mainka_sok:_2017,jonesoauth} trick an honest RP to connect to a malicious IdP following the issuance of an access token, and repeating authorization codes from the user to this malicious IdP.
\sysname uses a direct interaction between the user and the IdP, which is simpler to implement and reduces the potential for exploits.

A second category of attacks exploits the fact that in OIDC, a part of the communication between the IdP and an RP is relayed by the user browser using HTTP redirects.
Code/Token/State Leakage~\cite{jones2017oauth,fett_web_2017}, CSRF Attacks and Third-Party Login Initiation~\cite{fett2016comprehensive} are examples of exploits on incorrect OIDC implementations that do not properly check redirects or embed sensitive information such as ID tokens on redirection URLs.
307 Redirect Attack~\cite{fett2016comprehensive} similarly exploit the improper use of HTTP redirection codes.
\sysname only uses direct interactions between the user device and the IdP, in the setup phase, or an RP, in the sign-on phase, again reducing the risk of improper implementation and exploits.
The redirection by the RP to the IdP cached authentication page and Wasm client may trick users, but we rely on the security of the browser's password manager to mitigate this risk.


\vspace{-2mm}
\section{Evaluation}
\label{sec:evaluation}
\vspace{-1mm}

%

We evaluate the \sysname prototype and answer the following research questions:
\begin{enumerate}
 \itemsep0em
 \item Are \sysname costs and usage latencies adequate to replace OIDC as an SSO solution?
 How does \sysname performance compare to anonymous credential schemes providing similar security guarantees?
 \item Does the use of cryptographic operations at the user side impair the deployment of \sysname on low-power devices, such as mobile phones or tablets?
 \item What is the scalability of \sysname when using an increasing number of attributes in users' profiles?
 \item How do the implementations of the IdP and RP scale up when deployed in the cloud?
\end{enumerate}

\para{Setup}
We deploy an IdP and an RP on two \texttt{m5ad.xlarge} instances on Amazon EC2 (4 virtual cores, 16~GB of RAM each), both in the same EC2 region. 
We use two representative user devices: A Dell Latitude 5590 laptop with an Intel Core i7-8650U CPU and 16~GB of RAM, and a Raspberry~PI model 3b (RPI) with an A53 quad-core ARMv8 CPU and 1~GB of RAM.
The RPI is representative of the performance of lower-end mobile devices such as phones or tablets.
Both devices use Mozilla Firefox 76.0.1 to run the Wasm client.
User operations (\eg, entering a password) are emulated and instantaneous, to focus on the performance of the protocol.
We are open sourcing our implementation, benchmarking scripts, and measurements data to enable reproducible results
\footnote{
\ifdefined\cameraReady
\url{https://github.com/Zhiyi-Zhang/PSSignature}
\else
Link omitted for blind review.
\fi
}.

\para{Comparison to OIDC and IRMA}
We use as a first comparison point \texttt{pyoidc}~\cite{pyoidc}, a complete Python implementation of OIDC.
Note that the co-location of the RP and the IdP in the same EC2 zone also applies to the deployment of OIDC; this co-location is actually in favor of OIDC when measuring operation latencies.
We evaluate \texttt{pyoidc} with its default settings where a standard ID token is included in the AuthN response and a single attribute is retrieved by the RP.

Our second comparison point is IRMA~\cite{alpar2017irma}, an authentication system based on the Idemix anonymous credentials~\cite{idemix,camenisch2002design}.
IRMA is a state-of-the-art system in use by the Privacy by Design foundation~\cite{privacybydesign}.
We ported \texttt{irmago}, the implementation of IRMA for IoS/Android mobile platforms in \texttt{go}, to run on the same GNU/Linux platforms as \sysname.
We generate and deploy 4096-bit IRMA keys when issuing and verifying credentials\footnote{While the National Institute of Standards and Technology (NIST) allows 3072-bit keys until 2030, IRMA does not support this size. 4096-bit IRMA keys have security level equivalent to our implementation of \sysname.}.

All interactions in the three systems happen over \texttt{https}.

\begin{figure}[t]
	\centering
	\includegraphics[scale=0.34]{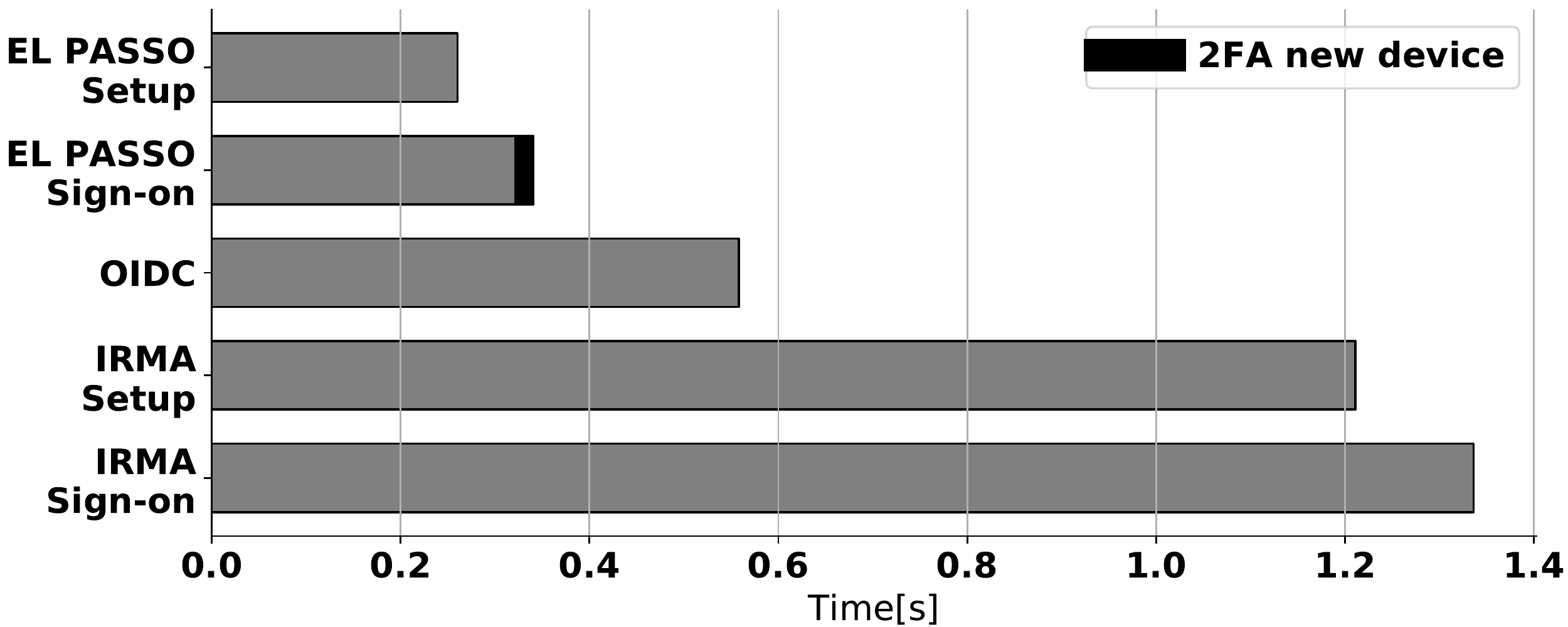}
	\caption{User-perceived operation latencies.}
	\label{fig:comparison_oidc_elpasso}
\end{figure}

\begin{figure}[t]
	\centering
	\includegraphics[scale=0.34]{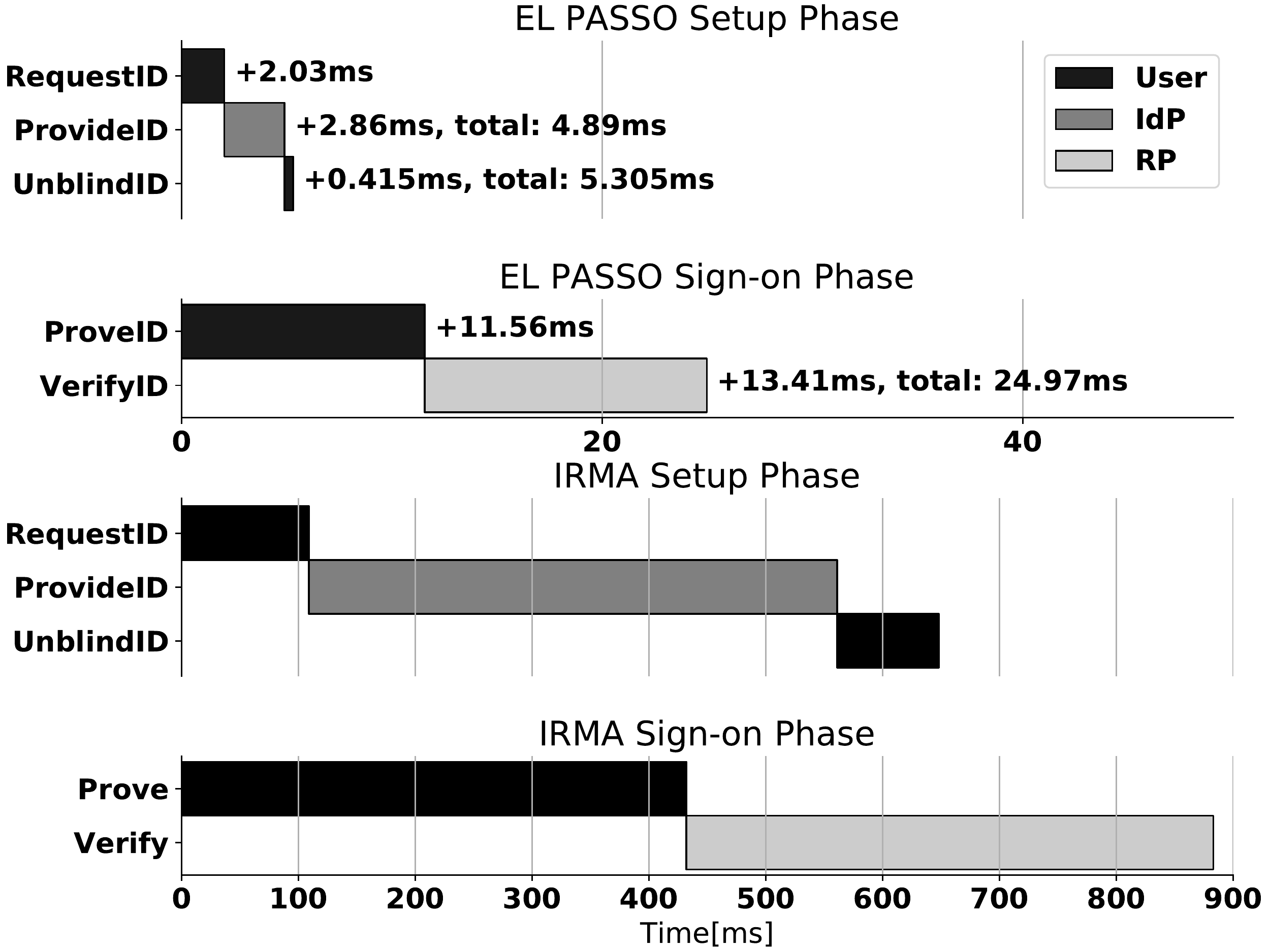}
	\caption{Breakdown of the execution time of computational phases in \sysname and IRMA.}
	\label{fig:gantt}
\end{figure}

\begin{figure} [t]
	\centering
	\includegraphics[width=\linewidth]{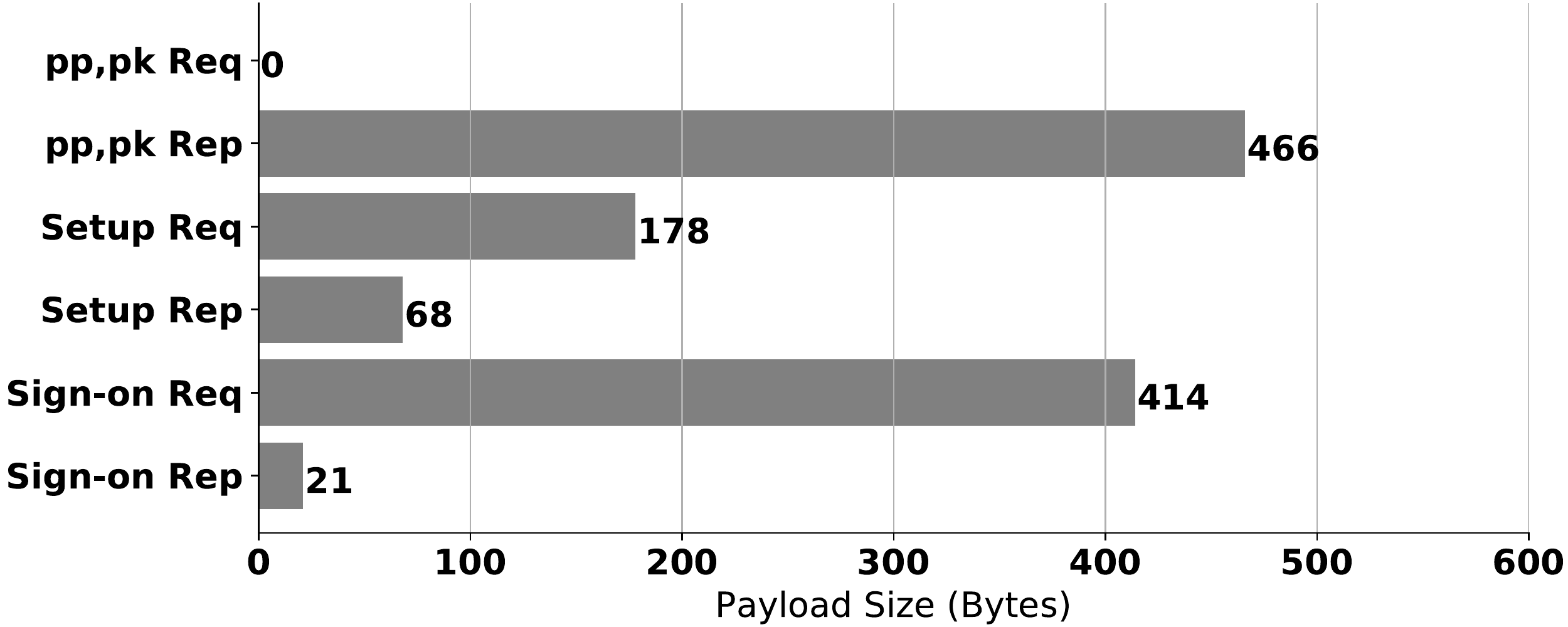}
	\caption{Payload size of messages exchanged in \sysname, for credentials with 3 attributes (first two lines are for the request of the public key of the IdP by the RP).}
    \label{fig:pkt-size}
\end{figure}

\para{Latency and costs}
We start with an evaluation of the latency of operations in \sysname, IRMA, and OIDC.
We use the laptop device, and credentials with the minimal number of 3 attributes $s$, $\gamma$, and $\mathit{tp}$.
We analyze the impact of changing the number of attributes in a later experiment.
\Cref{fig:comparison_oidc_elpasso} presents the complete latencies as perceived by the user.
These latencies include the latencies to and from the cloud, which we measured to be on average  20~ms round-trip.
We present also the breakdown of computational phases in the two protocols in \Cref{fig:gantt}, and the size of the payload of exchanged messages in \sysname in \Cref{fig:pkt-size}.

For \sysname and IRMA, we separate the asynchronous setup and sign-on phases, while sign-on in OIDC is a single, synchronous, and coupled operation.
We observe that authentication in OIDC takes less time than the two \sysname phases combined, in part because the RP and IdP are located in the same EC2 region--In most deployments, they would be deployed in different data centers, and the RP-IdP round-trip time would add to the overall latency.
However, the setup phase only takes place once per credential validity period, and in the majority of cases where credentials are already available at the user side, perceived latencies for sign-on will be \emph{lower} with \sysname than they are with OIDC.
IRMA experiences 5x higher Setup latency and 4x higher Sign-on latency, which is a result of much heavier cryptographic operations.
Associating a new device for 2FA during the sign-on phase results in only 10ms latency increase compared to a regular sign-on.

The breakdown of computational operations in \Cref{fig:gantt} allows identifying the CPU time required by the different phases (note that network latencies are not shown in the breakdown).
In contrast, \sysname requires little CPU time from the IdP, and only during the setup phase.
Overall, computational costs are slightly higher for \sysname, but they are also more decentralized, impacting mostly users and RPs.
A similar breakdown can be observed for IRMA.
However, the combined execution time is 100x higher for the setup phase and 39x higher for the sign-on phase.

The amount of payload exchanged, shown in \Cref{fig:pkt-size}, is reasonable.
The largest payload is the sign-on request from the client to the RP and is 0.5~KB in size.
We conclude this first set of experiments with a positive answer to our two first questions:
\sysname latencies and cost compare favorably to those of OIDC and would allow for deployment as an alternative SSO solution with negligible impact on performance or costs for users and operators of online services.
Furthermore, \sysname significantly reduces the user-perceived latency and computational time in comparison to a similar scheme based on anonymous credentials.

\para{Performance on low-power devices}
As the previous experiment has shown, \sysname requires computation and therefore CPU time at the user side.
We evaluate in this experiment whether these costs are acceptable for using it on low-power devices, such as mobile phones, tablets, or connected appliances.
Our setup is the same as with the previous experiment, but using the RPI device instead of the laptop.

\begin{table}[t!]
\footnotesize
\newcolumntype{C}{>{\raggedright\let\newline\\\arraybackslash\hspace{0pt}}m{0.28\linewidth} }
\newcolumntype{D}{>{\raggedright\arraybackslash} m{0.32\linewidth} }
\newcolumntype{E}{>{\raggedright\arraybackslash} m{0.32\linewidth} }
\resizebox{\linewidth}{!}{
\begin{tabular}{CDE}
\toprule
\textbf{Operation} & \textbf{Latency [s]} & \textbf{CPU time @ user [s]} \\ \midrule
EL PASSO Setup   & 0.72$\pm0.16$ (+190\%) & 0.11$\pm0.001$ (+397\%) \\
EL PASSO Sign-on  & 0.82$\pm0.18$ (+125\%) & 0.18$\pm0.004$ (+262\%) \\
OIDC             & 0.80$\pm0.02$ (+45\%) & NA \\
IRMA Setup & 30.295$\pm0.39$ (+2420\%) & 29.68$\pm0.27$ (+4390\%) \\
IRMA Sign on & 34.182$\pm0.49$(+2458\%) & 33.891$\pm0.43$ (+3640\%) \\
\bottomrule
\end{tabular}
}
\caption{\sysname performance using a Raspberry~PI for single and multi (M) device scenario, relative to results using a laptop from Figures~\ref{fig:comparison_oidc_elpasso} and \ref{fig:gantt}.}
\label{tab:pi_timing}
\end{table}

\begin{figure*}[t!]
	\centering
	\includegraphics[scale=0.34]{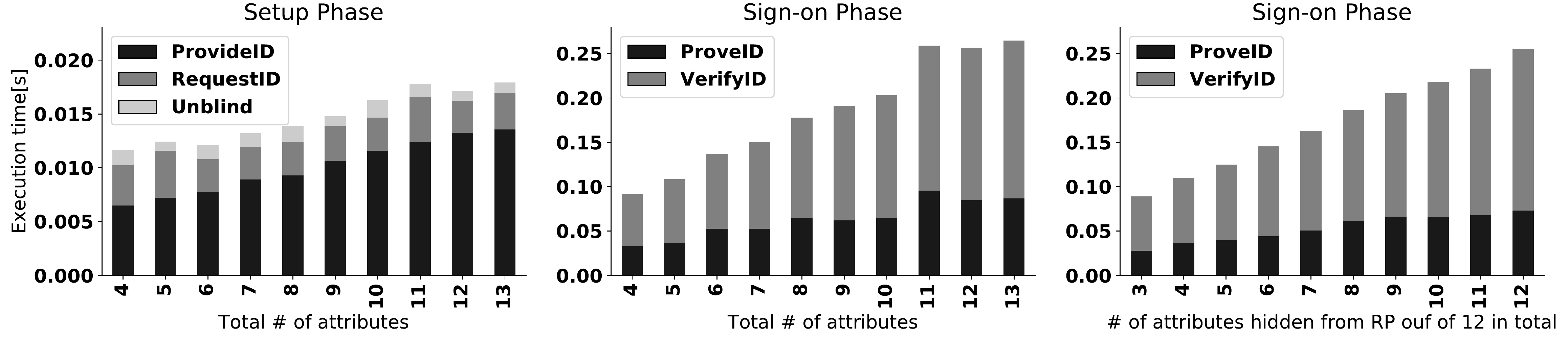}
	\caption{Impact of the the total number of attributes and the number of attributes hidden from the RP on CPU time.}
	\label{fig:attributes}
\end{figure*}

\Cref{tab:pi_timing} compares the perceived latency using the RPI to those in \Cref{fig:comparison_oidc_elpasso}, and the total CPU time at the user side, compared to \Cref{fig:gantt}.
We can observe that the CPU cost for the setup phase almost quadruples, yet remains low at 110~ms.
For the sign-on phase, the cost is multiplied by 4, primarily due to the lower performance of cryptographic operations on the ARM CPU.
Yet again, the overall CPU time remains within acceptable boundaries at less than 200~ms and 220~ms when adding a new device to an account.
The overall latency is impacted by both this increase in CPU time (except for OIDC), and the performance of the browser running on the RPI (including for OIDC).
All operations succeed in a reasonable time, the longest being the sign-on taking a second on average, only slightly higher than OIDC compared to the previous experiment.
In contrast, more complex IRMA operations experience significant execution time increase and result in Setup and Sign-on phase finishing in more than 30s.
This allows us to answer positively to our second question: The performance and costs of \sysname make it adequate as a solution for SSO, even when users are equipped with low-power or mobile devices.

\para{Scalability in the number of attributes}
We investigate the impact of the number of attributes embedded in user credentials on the computational cost of \sysname.
The two first plots of \Cref{fig:attributes} show the evolution of CPU time with a growing number of attributes all of which are hidden from the RP.
Note that the case of 3 attributes corresponds to the data in \Cref{fig:gantt}.
As expected, the CPU time increases linearly for both the setup phase and sign-on phase (first and the second plot, respectively).
This increase is primarily due to the additional complexity of the \textsf{ProveID} operation, due to the need to respectively create and validate zero-knowledge proofs for more values.
Yet, the total cost, even with 13 attributes, remains reasonable, at less than a second of total CPU time.
The third plot evaluates the cost of the sign-on phase when the user decides to hide an increasing number attributes from the RP, from a profile with 12 attributes: An abscissa value of 9 means, therefore, that the preparation of the credential for this RP only reveals 3 attributes\footnote{Our design requires at least 2 attributes ($s$, $\gamma$) to be hidden from RPs}.
As expected, hiding more attributes increases the computational load in the ProveID and VerifyID parts of the algorithm, yet again requiring less than a second of total CPU time.
We conclude, therefore, that \sysname scales sufficiently well with the number of attributes to be used in practical scenarios, where the identity of a user is formed of up to a dozen different fields, answering our third question.

\begin{figure}[t!]
	\centering
	\includegraphics[scale=0.34]{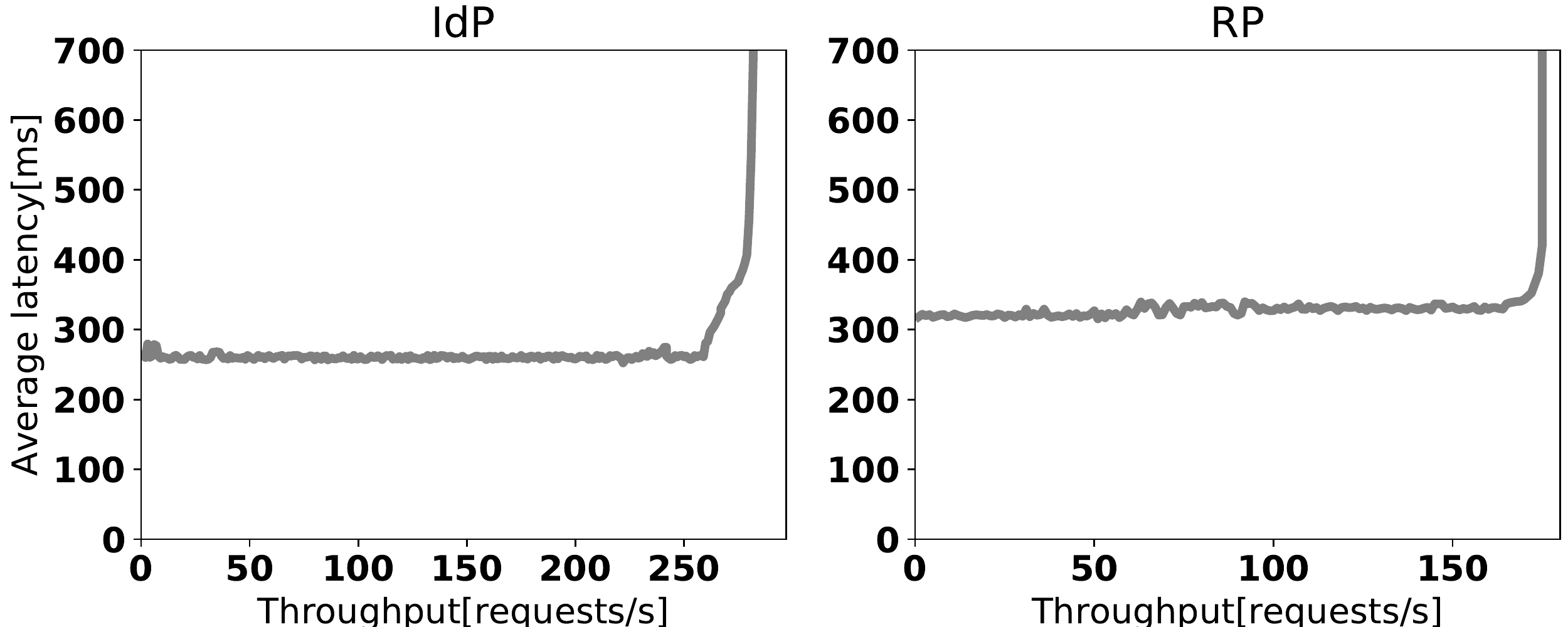}
	\caption{Scaling of \sysname services in the cloud.}
	\label{fig:throughput}
\end{figure}

\para{Scalability of the IdP and RP}
In this last experiment, we measure the scalability of the \sysname implementation in the cloud, for a large number of users.
We inject a growing number of precomputed requests in parallel from the laptop device and measure the achieved throughput and operation latencies.
\Cref{fig:throughput} is a parametric plot showing the relation between the two measurements.
The simpler operations required by the setup phase allow a single IdP node to handle up to 272 requests per second.
The costlier sign-on phase at the RP lowers the number of operations per second to about 169\footnote{We note that operations at the IdP and RP for different users are naturally disjoint-access parallel, if the user information is stored in a scalable NoSQL database.
This allows scaling the IdP or RP horizontally as necessary.}.
These final measurements prove that \sysname, while involving privacy-preserving mechanisms can still be easily deployed on commodity cloud servers, and positively answer our fourth question.


\vspace{-2mm}
\section{Related Work}
\label{sec:related}
\definecolor{__red}{rgb}{0.8,0.1,0.1}
\definecolor{__green}{rgb}{0.1,0.6,0.1}

\newcommand{\yestick}{{\color{__green}{\cmark}}}
\newcommand{\notick}{{\color{__red}{\xmark}}}

\begin{table*}[t]
\centering
\resizebox{\textwidth}{!}{%
\begin{tabular}{  c | c c c c c c c c c c}
\toprule
\bf System & \bf \thead{Personal\\Authentication} & \bf \thead{Intra-RP\\Linkability} & \bf \thead{Tracking\\Protection} & \bf \thead{Selective\\Disclosure} & \bf \thead{Inter-RP\\Unlinkability} & \bf \thead{Provable Personal\\Attributes} & \bf \thead{Reliable Identity\\Retrieval} & \bf \thead{Asynchronous\\Authentication} & \bf \thead{No RP\\registration }& \bf \thead{Browser-only} \\
\midrule
SAML~\cite{hughes2005security} & \notick  & \yestick & \L & \L & \notick & \notick & \yestick & \notick & \notick & \yestick \\
OpenID Connect~\cite{openid} & \notick & \yestick & \L & \M & \notick & \yestick & \yestick & \notick & \notick & \yestick\\
Apple SignIn~\cite{sia} & \notick & \yestick & \L & \M & \yestick & \yestick & \yestick & \notick & \notick & \yestick \\
SPRESSO~\cite{fett2015spresso} & \yestick & \yestick  & \H & \M & \notick & \yestick & \yestick & \notick & \notick & \yestick \\
PRIMA~\cite{asghar2018prima} & \yestick & \yestick & \H & \M & \notick & \yestick & \yestick & \notick & \yestick & \notick \\
\midrule
UnlimitID~\cite{isaakidis2016unlimitid} & \yestick & \yestick  & \M & \H & \yestick & \yestick & \notick & \notick & \notick & \notick \\
NextLeap~\cite{halpin2017nextleap} & \yestick & \yestick  & \M & \H & \yestick & \yestick & \notick & \notick & \yestick & \notick \\
UProve~\cite{paquin2011u, paquin2011uc} & \yestick & \notick & \H & \H & \yestick& \yestick & \notick & \yestick & \yestick & \notick \\
Privacy-ABCs~\cite{rannenberg2015attribute} & \yestick & \yestick  & \H & \H & \yestick & \yestick & \yestick & \yestick & \yestick & \notick \\
IRMA~\cite{alpar2017irma} & \yestick & \notick  & \H & \H & \yestick & \yestick & \notick & \yestick & \yestick & \notick \\
Hyperledger Idemix~\cite{alpar2017irma} & \yestick & \notick  & \H & \H & \yestick & \yestick & \notick & \yestick & \yestick & \notick \\
\midrule
\rowcolor{verylightgray}
\textbf{\sysname} & \yestick & \yestick & \H & \H & \yestick & \yestick & \yestick & \yestick & \yestick & \yestick \\
\bottomrule
\end{tabular}}
\caption{Properties of different SSO and Anonymous Credentials systems.
\footnotesize
\\ \textbf{Tracking Protection} \M $\colon$ UnlimitID and UnlimitID-based NextLeap rely on unlinkable credentials. However, the blinded credentials must be deposit by the users at IdP, potentially allowing IdP to perform user tracking.
--- \textbf{Selective Disclosure} \M $\colon$ OIDC, Apple SignIn and SPRESSO allow to disclose a subset of user information, but are unable to prove statements about their attributes (\ie age > 18). PRIMA supports proving statement about attributes only if they are expressed as additional attributes signed by IdP.
}
\label{tab:related}
\end{table*}

We review related work on SSO, its privacy-preserving extensions and anonymous authentication.
We classify the most related of the systems we discuss in \Cref{tab:related} using the properties defined in \Cref{sec:goals} and summarized in \Cref{tab:sso_properties}.

\para{SSO Standards}
The Security Assertion Markup Language (SAML)~\cite{hughes2005security} is an XML-based authentication protocol, widely deployed before OIDC was standardized.
It uses a message flow that is very similar to that of OIDC, therefore shares its privacy vulnerabilities.
Furthermore, SAML does not enable selective attribute disclosure and provides less flexibility to developers than OIDC.

OIDC combines the previous Open~ID identity management standard with the OAuth authentication protocol~\cite{hardt2012oauth}.
The privacy issues of these protocols were pointed out as being a result of the direct IdP and RP communication~\cite{openid-problems}.

%

\para{SSO extensions}
Sign In with Apple~\cite{sia} uses randomized per-RP identifiers (alias email addresses) for users instead of permanent identifiers (actual email address).
This solution provides inter-domain unlinkability.
However, Apple has largely adopted OIDC for its implementation~\cite{openid-sia,sia} and the IdP-side privacy concerns also hold true for this system.

SPRESSO~\cite{fett2015spresso} decouples the communication between the RP and the IdP, letting the two parties communicate indirectly through a \emph{forwarder} agent at the client.
A user sign-on request to an RP is followed by a synchronous user request to the IdP for credentials.
The synchronicity of operations requires protection against time-based attacks, where the IdP could correlate requests from the user and the RP.
Furthermore, SPRESSO leaks user's global ID to RPs enabling tracking

PRIMA~\cite{asghar2018prima} decouples communications between RP and IdP and supports selective attributes disclosure on top of Oblivion~\cite{simeonovski2015oblivion}.
However, it requires contacting the IdP for every user sign-on and does not provide inter-RP unlinkability.

\para{Anonymous authentication}
Anonymous credentials such as CL Signatures~\cite{cl,lee2013aggregating} and Idemix~\cite{idemix,camenisch2002design} are useful in personal identity management~\cite{alpar2017irma}, anonymous attestation~\cite{brickell2004direct, chen2010design, bernhard2013anonymous}, and electronic cash~\cite{canard2015divisible}.
They provide blind issuance and unlinkability through randomization, but come with significant computational overheads, and large credentials size.
U-Prove~\cite{paquin2011u, paquin2011uc} and Anonymous Credentials Light (ACL)~\cite{acl} are computationally efficient credentials that can be used once unlinkably; therefore the size of the credentials is linear with the number of unlinkable uses.
Furthermore, they do not allow an RP to distinguish different sign-on attempts by the same user, and cannot provide intra-RP linkability.
UnlimitID~\cite{isaakidis2016unlimitid} builds attribute-based SSO credentials over aMAC~\cite{amac}, used as pseudonyms.
This allows inter-RP unlinkability, as IdPs are unable to track user activity over different RPs using different pseudonyms.
UnlimitID follows the main flow of OIDC and requires users to deposit their anonymized pseudonyms at IdP before RPs can access them.
This may allow the IdP to correlate the deposit of a pseudonym and its request by an RP, enabling tracking.
The NextLeap project~\cite{halpin2017nextleap} intends to extend UnlimitID~\cite{isaakidis2016unlimitid} by storing identity and trust information in a blockchain, positioning that this would remove the need for RPs to explicitly register with IdPs, as is the case in \sysname.
Recent attribute-based credential~\cite{camenisch2013concepts} implementations such as IRMA~\cite{alpar2017irma}, Privacy-ABCs~\cite{rannenberg2015attribute,abcimplem,ibmidemix} and Hyperledger Idemix~\cite{idemixgen,ursa} significantly improve the performance, but still suffers from high user-perceived latency on less powerful devices. Furthermore, they require manual installation and configuration/credential management, do not enable 2FA or multi-device support. 
Similar issues were already identified as barriers preventing wide-spread deployment of mature security systems such as PGP~\cite{ruoti2013confused,whitten1999johnny}.

In combination with anonymous credentials, multiple works propose to prevent~\cite{tsang2007blacklistable, camenisch2006win, brands2007practical,camenisch2001efficient} or limit~\cite{henry2011formalizing} login attempts of specific users without revealing their identities.
While those platforms can block misbehaving users from accessing a specific RP, they are unable to hold these users accountable for their actions (\eg when publishing hate speeches online).
Finally, these blacklisting systems require significant computational and communication overhead, limiting their usability and deployability, which are essential goals for \sysname.



\section{Conclusion}
\label{sec:conclusion}

We presented \sysname, an SSO solution that combines the security of anonymous credentials with the practicality of OIDC.
Our solution protects users from being tracked by either RPs or IdPs and allows us to disclose only the minimum user information required to sign on.
While providing strong privacy protection, \sysname can also hold misbehaving users accountable in cooperation with law enforcement authorities.
Our system is implemented as a Wasm module that is downloaded on the fly and cached by the user's browser. 
Support for multi-device deployment, privacy-preserving 2FA, and device theft recovery is provided and only rely on the user browser's built-in features. 
We believe that these properties open the perspective of using our system in a wide range of use cases where the use of anonymous credentials would otherwise be an issue, such as e-democracy platforms and opinion forums.
%
%

\ifdefined\cameraReady
\section*{Acknowledgments}
This work is partially supported by  the the National Science Foundation under award CNS-1629922, the Belgian FNRS project DAPOCA (33694591), and Facebook Calibra.
The authors would like to thank Dahlia Malkhi and Ben Maurer for their feedback on an earlier version of this work.
\fi

\bibliographystyle{plain}
\bibliography{references}

\end{document}